\documentstyle[12pt,epsf]{article}
\newcommand{\bn}{{\bf n}}
\newcommand{\MM}{{\cal M}}
\newcommand{\A}{{\cal A}}
\newcommand{\B}{{\bf\cal B}}
\newcommand{\cd}{\cdot}

\newcommand{\de}{\delta}
\newcommand{\De}{\Delta}

\newcommand{\ga}{\gamma}
\newcommand{\Ga}{\Gamma}

\newcommand{\la}{\lambda}
\newcommand{\Om}{\Omega}
\newcommand{\om}{\omega}
\newcommand{\si}{\sigma}

\newcommand{\ra}{\rightarrow}

\newcommand{\lap}{\triangle}

\newcommand{\bm}[1]{\mbox{\boldmath $#1$}}
\newcommand{\be}{\begin{equation}}
\newcommand{\ee}{\end{equation}}

\newcommand{\lsim}{\stackrel{<}{\sim}}
\newcommand{\bea}{\begin{eqnarray}}
\newcommand{\eea}{\end{eqnarray}}
\newcommand{\bean}{\begin{eqnarray*}}
\newcommand{\eean}{\end{eqnarray*}}
\newcommand{\dd}{\partial}

\newcommand{\bk}{{\bf k}}
\newcommand{\bx}{{\bf x}}

\begin{document}          
\begin{center}
{\Large\bf Microwave Background Anisotropies \\[0.2cm]
from Alfv\'en waves}\\[0.5cm]
{\large R. Durrer$^*$, T. Kahniashvili$^\dagger$ and A. Yates$^*$ 
	}\\[0.2cm]
 $^*$D\'epartement de Physique Th\'eorique, Universit\'e de Gen\`eve,\\
24 quai Ernest Ansermet, CH-1211 Gen\`eve 4, Switzerland\\
$^\dagger$Department of Astrophysics, Abastumani 
Astrophysical Observatory,\\
Kazbegi Ave. 2a,  380060 Tbilisi, Georgia\\
\end{center}
\begin{abstract}
We investigate microwave background anisotropies in the presence of
primordial magnetic fields. 
We show that a homogeneous  field 
with fixed direction can amplify vector perturbations.
We calculate the correlations of $\delta T/T$ explicitly and show that
 a large scale coherent field induces correlations between
$a_{\ell-1,m}$ and $a_{\ell+1,m}$.
We discuss constraints on the amplitude and spectrum
of a primordial magnetic field imposed by observations of CMB
anisotropies. 
\end{abstract}
\section{Introduction}
Since the detection of CMB anisotropies by the COBE satellite\cite{COBE},
 it has become clear that
anisotropies in the cosmic microwave background (CMB) provide a 
powerful tool for distinguishing models of 
cosmological structure formation. Furthermore, they may help to determine
cosmological parameters which influence their spectrum in a well defined,
non-trivial way \cite{berkeley}. It is 
thus important to calculate the CMB anisotropies for a given
model simply and reliably.  

The origin of observed galactic magnetic fields of the order of
 $\mu$Gauss is still unknown. For some time it has been
believed that tiny seed fields can be amplified by a non-linear galactic dynamo
mechanism. The effectiveness of this process has recently been
strongly questioned, however \cite{critics}. If magnetic fields are
not substantially amplified by non-linear effects, but have just
contracted with the cosmic plasma during galaxy formation, primordial 
fields of the order $10^{-9}$ Gauss on Mega-parsec scales are required
to induce the observed galactic fields.

It is interesting to note that a field strength of 
$10^{-8}$Gauss provides an energy density of $\Om_B = B^2/(8\pi\rho_c) \sim
10^{-5}\Om_{\ga}$, where $\Om_{\ga}$ is the density parameter in
photons. We naively
 expect a  field of this amplitude to induce perturbations
in the CMB on the order of $10^{-5}$, which is just the level of the
observed anisotropies. This leads us to investigate the extent to
which the isotropy of the CMB may constrain primordial magnetic fields.
It is clear from our order of magnitude estimate that we shall never be
able to constrain tiny seed fields on the order of $10^{-13}$Gauss with this
approach, but primordial fields of $10^{-9}$ to $10^{-8}$Gauss may well have
left observable traces in the microwave background.

This is the question we investigate quantitatively in this paper.
Some work on the influence of primordial magnetic fields on CMB
anisotropies has already
been published, particularly the cases of fast magneto-sonic waves and
the gravitational effects of a constant magnetic field
\cite{Ad,BFS,Barrow}. Here we study Alfv\'en waves. 
We leave aside the problem of generation of primordial magnetic fields.
This issue is addressed {\em e.g.} in \cite{Veneziano,GS}.

The possible influence of magnetic fields on large scale structure 
formation has been investigated in \cite{Olinto} and references therein. 

The paper is organized as follows.
In Section~2 we discuss the effect of a homogeneous magnetic field
background on the cosmic plasma.
Both scalar (potential, or fast and slow magneto-sonic waves) and
vorticity  (Alfv\'en) waves can be induced. We study the latter.
In Section ~3 we consider
the influence of  Alfv\'en waves on CMB anisotropies.  The influence of the
magneto-sonic waves can be interpreted as a slight change in the speed
 of sound, and
has been investigated in \cite{Ad}. In Section~4 we present our conclusions.
Some of the more technical computations as well as a discussion of Silk
damping of vector perturbations are left to two appendices.

{\bf Notation:} For simplicity we concentrate on the case
$\Om_0=1$. The choice of $\Om$ is of little importance for our
perturbation variables (which have to be calculated for 
at early times, when curvature effects
are not significant), but does influence the
resulting $C_\ell$'s due to projection effects.
Throughout, we use conformal
time which we denote by $t$. The unperturbed metric is thus given
by $ds^2=a^2(t)(-dt^2+\de_{ij}dx^idx^j)$.
Greek indices run from $0$ to $3$, Latin ones from $1$ to $3$.
We denote spatial (3d) vectors with bold face symbols. 

\section{Cosmological vector perturbations and Alfv\'en waves}

Vector perturbations of the geometry are of the form
\be
	\left(h_{\mu\nu}\right) =\left(\begin{array}{cc}
	0 & B_i \\
	B_j & H_{i,j} + H_{j,i}\end{array} \right)~,
\label{gvec} 
\ee
where  ${\bf B}$ and ${\bf H}$ are divergence-free, 3d vector fields
supposed to 
vanish at infinity. Studying the behaviour of these variables under
infinitesimal coordinate transformations (called gauge transformations
in the context of linearized gravity), one finds that the
combination
\be
 \bm{\si} = \dot{\bf H}-{\bf B} \label{siv}
\ee
is gauge invariant. Geometrically, $\bm{\si}$ determines the 
vector contribution
to the perturbation of the extrinsic curvature \cite{KS,fc}.

To investigate perturbations of the energy-momentum
tensor, we consider a baryon, radiation and cold dark matter (CDM)
universe for which anisotropic stresses are negligible. The only vector
perturbation in the energy-momentum tensor is thus a
perturbation of the energy flux, $u$, the time-like eigenvector of
$T_\mu^{~\nu}$. We parameterize a vector perturbation of $u$ with a
divergence free vector field $\bf v$, such that
\be
{\bf u} = {1\over a}{\bf v} ~.
\ee 
Analyzing the gauge transformation properties of $\bf v$, one finds
two simple gauge-invariant combinations \cite{KS},
\be
 {\bf V} = {\bf v-\dot{H}}  ~~~\mbox{ and }~~~~
  {\bf \Om} ={\bf v-B} ~.
\ee
They are simply related by
\be {\bf V = \Om} -\bm{\si} ~. \label{rel}
\ee
The perturbations of the Einstein equations, together with energy-momentum
conservation, yield \cite{fc}
\bea
-{1\over 2}\lap\bm{\si} &=&
3\left( {{\dot a} \over a} \right)^2
{\bf \Om} ~,\label{const}\\
 \dot{\bm{\si}} + 2\left({\dot{a}\over a}\right)\bm{\si}&=& 0~,
\label{dyn} \\ 
\dot{\bf\Om} + (1-3c_s^2){\dot{a}\over a} {\bf \Om} &=& 0~. \label{cons}
\eea
The two Einstein equations~(\ref{const}, \ref{dyn}) and the
momentum conservation equation (\ref{cons}) are not
independent.  Eq.~ (\ref{cons}) follows
from Eqs.~(\ref{const}) and (\ref{dyn}).

This system does clearly not describe waves.
From Eq.~(\ref{dyn}) it follows that $\bm{\si}$ decays like
$1/a^2$. Furthermore, Eq.~(\ref{const}) implies $\bm{\Om} \propto
(kt)^2\bm{\si}$. In the radiation dominated era, where $a\propto t$,
this yields $\bm{\Om} ={1\over 6} (kt_{in})^2\bm{\si}_{in}$, where $t_{in}$
is some initial time at which fluctuations were created, {\it e.g.} the
end of inflation. The fact that $\bm{\Om}$ stays constant during the radiation
dominated era also follows from Eq.~(\ref{cons}). 
On cosmologically interesting scales, $k\ll
1/t_{in}$, we have therefore $\bm{\Om}\ll\bm{\si}_{in}$ and 
$\bm{\si}\ll\bm{\si}_{in}$. In contrast, scalar and tensor
perturbations remain constant on super-horizon scales. For this
reason, vector perturbations induced at a very early
epoch (e.g. inflation) which have evolved freely can be entirely
neglected in comparison to their scalar and tensor counterparts.

This situation is altered in the presence of a primordial magnetic field,
which induces vorticity waves after the inflationary era.
Let us consider a homogeneous magnetic field 
${\bf B}_0$ before the time of decoupling of matter and radiation. 
Such a field could have originated, 
for example, at the electroweak phase transition \cite{GS}.
When the photon-baryon fluid is taken to be a perfectly conducting
plasma, an external magnetic field induces two distinct modes of
oscillation. Magneto-sonic waves, scalar perturbations which 
propagate at speeds $c_{\pm}$ slightly above or very much below the
ordinary speed of sound in the plasma. These induce
density oscillations just like ordinary acoustic waves. The slight
change in the speed of sound can change slightly the position and shape 
of the acoustic
peaks in the CMB spectrum \cite{Ad}. Here we discuss the vectorial
{\it Alfv\'en waves}.

We assume a plasma with infinite conductivity and use the  
frozen-in condition: 
\[
\bf E +   v \times  B = 0~,
\] 
where ${\bf v}$ is the plasma velocity field. Our  
plasma is non-relativistic ($v \ll 1$). The field lines of a
homogeneous background magnetic field in a Friedmann universe are 
just conformally diluted, such that $B_0\propto 1/a^2$.  Until
recombination, the photon-baryon plasma is dominated by photons,
$\rho_{r} \simeq \rho_{\gamma}  \propto a^{-4}$ ($\rho_{r}$ denotes the
 combined baryon and photon energy density) and 
the ratio $B_0^2/(\rho_r+p_r)$ is time-independent.  
We study purely vortical waves which induce a
fluid vorticity field $\bm{\Omega}(\bk)$  normal to $\bk$. 
We note that charged particles are tightly coupled to the
radiation fluid and obey the equation of state
$p_r = \rho_r /3$. 

It is convenient to rescale physical quantities like the fields and
the current density as follows: 
\[
    {\bf E} \ra{\bf E}a^2 ~,~~{\bf B}\ra {\bf B} a^2~~ 
\mbox{ and }~~~{\bf J} \ra{\bf J} a^3~.
\]
We now introduce first-order vector perturbations
in the magnetic field  (${\bf B}_1$) and in the fluid velocity
($\bm{\Omega}$);
\be
	{\bf B}={\bf B}_0 +{\bf B}_1, ~~~
	\bm{\nabla}\cd{\bf B}_1=0 ~~~ \mbox{ and}\label{ans1}
\ee
\be
	{\bf v}= \bm{\Omega}~,~~~~
	\bm{\nabla}\cd\bm{\Omega}=0  ~.\label{ans2}
\ee
To obtain the equations of motion for $\bm{\Omega}$ and ${\bf B}_1$,
we first
consider Maxwell's equations. Since the fluid velocity is small, we
may neglect the displacement current in Amp\`ere's law, which then
yields
\be
{\bf J} ={1 \over 4\pi}\bf{ \nabla\times} {\bf B_1 }. \label{Amp} 
\ee
We replace ${\bf E}_1$ with ${\bf B}_0$, using the frozen-in
condition. The induction law then gives
\be
{\partial  \over \partial t} {\bf B}_1= {\bf  \nabla} \times( 
{\bf v }\times {\bf B}_0)~.  \label{ind} 
\ee
Inserting relation~(\ref{Amp}) for  the current, 
the equation of motion ($T^{i\mu}_{;\mu}=F^{i\mu}j_\mu$) 
for vector perturbations in the plasma becomes
\be
{\partial \over \partial t} {\bf v} =  - {1 \over 4 \pi (\rho_r+p_r)} 
{\bf  B}_0 \times ( {\bf \nabla} \times {\bf B}_1). 
\ee
(We have neglected viscosity, which is a good
approximation on scales much larger than the Silk damping scale \cite{Silk}.)
Taking the time derivative of this equation, we obtain with the help
of Eqs.~(\ref{ans1}), (\ref{ans2}) and (\ref{ind}) for a fixed Fourier
mode {\bf k}
\be
\ddot{\bm{\Omega}} ={({\bf B}_0\cdot \bk)^2 \over 4\pi(\rho_r+p_r)}
	\bm{\Omega}~~ \mbox{ and} \label{vort}
\ee
\be
\dot{\bm{\Omega}} ={i{\bf B}_0\cdot \bk \over 4\pi(\rho_r+p_r)}
	{\bf B}_1~. \label{B1}
\ee
These equations\footnote{Our derivation is valid either in a gauge
invariant framework as outlined in \cite{HPA} or in a gauge with 
vanishing shift vector. In other gauges metric coefficients will 
enter and complicate the equations.}
describe waves propagating at the velocity 
$v_A({\bf e}\cdot\hat{\bk})$, where
\[
v_A^2 = {B_0^2 \over 4 \pi (\rho_r+p_r)} ~,~~~ v_A\sim
4\times 10^{-4}(B_0/10^{-9}\mbox{Gauss})
\]
is the Alfv\'en velocity 
and ${\bf e}$ is the unit vector in the direction of the magnetic
field. Typically the Alfv\'en velocity will be
very much smaller than the speed of acoustic oscillations in the
radiation-dominated plasma ($c_s^2 = 1/3  \gg v_A^2$).

Due to the
observed isotropy of the CMB, we have to constrain
the magnetic field contribution to the total energy density. For
example, in the radiation dominated era it must be a fraction of less than
about $10^{-5}$ \cite{BFS}, leading to $v_A\lsim 10^{-3}$. 
Eq.~(\ref{vort}) is homogeneous in $\bm{\Omega}$ and so does not
determine the amplitude of the induced vorticity. The general
solution contains two modes, $\cos (v_A kt 
\mu)$ and $ \sin (v_A kt \mu)$ (where 
$\mu = {\bm{e}} \cdot \hat \bk$). If the cosine mode is present, it
dominates on the relevant scales $k<1/(v_At_{dec})$.  Then we 
can approximate $\cos (v_A kt_{dec}\mu) \simeq 1$ and the sine is
negligibly small. But this mode then describes the usual vector
perturbations without a magnetic field. We assume it to be absent.
We want to consider initial conditions, then, with
\[
\bm{\Omega}(\bk, t=0) = 0.
\]
Only the sine mode is present and we have
\be
\bm{\Omega}(\bk, t) = \bm{\Om}_0 \sin (v_A kt \mu) \simeq  
	\bm{\Om}_0 v_A kt \mu~.
\label{mag_vel}
\ee
The initial amplitude of $\bm{\Omega}_0$ is connected with the amplitude 
of ${\bf B}_1$ by means of Eq.~(\ref{B1}), yielding
\be
|\bm{\Omega_0}|=(v_A/B_0)|{\bf B}_1| ~. \label{amp}
\ee
This allows a vorticity amplitude of up to the order of the
Alv\'en velocity (see also \cite{Barrow}).

\section{CMB anisotropies from Alfv\'en waves}
Vector perturbations induce anisotropies in the CMB via a
Doppler effect and an integrated Sachs-Wolfe term \cite{fc}
 \be
\left({\De T\over T}\right)^{(vec)}= 
	-\left.{\bf V\cd n}\right|_{t_{dec}}^{t_0} +
  \int_{t_{dec}}^{t_0}\dot{\bm{\si}}\cd {\bf n}d\la   ~, \label{Tv} 
\ee
where the subscripts $dec$ and $0$ denote the decoupling epoch
($z_{dec}\sim 1100$) and today respectively. 
Since the geometric perturbation $\bm{\si}$ is decaying, the integrated
term is dominated by its lower boundary and just cancels $\bm{\si}$ in
${\bf V} =\bm{\Om-\si}$.
Neglecting a possible dipole contribution from vector perturbations 
today, we obtain
\be
{\delta T \over T} (\bn, \bk) \simeq \bn\cdot\bm{\Om}({\bf
k},t_{dec})=  \bn\cdot\bm{\Om}_0 v_A kt_{dec} ({\bf e}
\cdot \widehat \bk). 
\label{tb}
\ee
We assume that the vector perturbations $\bm{\Om_0}$ are created by some
isotropic random process, and so have a power spectrum of the form
\be
	\left\langle\Om_{0i}({\bf k})\Om_{0j}({\bf k})\right\rangle
	=(\de_{ij}-\hat{k}_i\hat{k}_j)A(|{\bf k}|) \label{spec}~.
\ee
For simplicity, we further assume that the spectrum 
$A(k)=(1/2)|{\bf {\Omega}_0}|^2(k)$ is a simple power law over the 
range of scales relevant here,
\be
A(k)=A_0 {k^n\over k_0^{(n+3)}}, \qquad k < k_0, \label{A_spec}
\ee
for some dimensionless constant $A_0$ and cutoff wavenumber $k_0$.
With this we can calculate the CMB anisotropy spectrum.

The $C_\ell$'s are defined by
\be
 \left.\left\langle{\delta T\over T}({\bf n}){\delta T\over T}({\bf n}')
\right\rangle\right|_{{~}_{\!\!(\bn \cdot \bn'=\mu)}} =
  {1\over 4\pi}\sum_\ell(2\ell+1)C_\ell P_\ell(\mu)~. 
\label{defCl}
\ee
A homogeneous magnetic field induces a preferred direction ${\bf e}$
and the correlation function (\ref{defCl}) is no longer a function of
$\bn \cdot \bn'=\mu$ alone but depends also on the angles between $\bn$
and  ${\bf B}_0$ as well as  $\bn'$ and  ${\bf B}_0$.  Statistical 
isotropy is broken. Setting
\be 
  {\delta T\over T}({\bf n})=\sum_{\ell,m}a_{\ell m}Y_{\ell m}(\bn),
\ee
in the isotropic situation, the $C_\ell$'s of Eq.~(\ref{defCl}) are just
\be
C_\ell = \langle a^{}_{\ell m}a^*_{\ell m}\rangle~, \label{Cl}
\ee
where $\langle \rangle$ denotes a theoretical expectation value over
an ensemble of statically identical universes.
We find that the presence of the preferred direction ${\bf B}_0$ not only
leads to an $m$-dependence of the correlators,  but also induces
correlations between the multipole amplitudes $a_{\ell+1, m}$ and
$a_{\ell-1, m}$. 
Correlations in the temperature fluctuations at different points on
the sky are no longer simply functions of their relative angular
separation, but also of their orientations with respect to the external
field. Detailed computations of the correlators for the Doppler
contribution from Alfv\'en waves are presented in Appendix~A. We obtain
\bea 
C_\ell(m) &=& \langle a^{}_{\ell m}a^*_{\ell m}\rangle  \nonumber\\
	&=& A_0v_a^2\left({t_{dec}\over
t_0}\right)^2(k_0t_0)^{-(n+3)}{2^{n+1}\Ga(-n-1)\over
\Ga(-n/2)^2} \times \nonumber\\
&& \!\!\!\!\!\!\!\!\!\!
\left({2\ell^4+4\ell^3-\ell^2-3\ell+6m^2-2\ell m^2-2\ell^2 m^2\over
(2\ell-1)(2\ell+3)}\right){\Ga(\ell+n/2+3/2)\over \Ga(\ell-n/2+1/2)} \\
D_{\ell}(m) &=& \langle a^{}_{\ell-1, m}a^*_{\ell+1, m}\rangle =
\langle a^{}_{\ell+1, m}a^*_{\ell-1, m}\rangle \nonumber\\
&=&  A_0 v_A^2\left({t_{dec}\over
t_0}\right)^2(k_0t_0)^{-(n+3)}{2^{n+2}\Ga(-n-1)\over |n+1|
\Ga(-(n+1)/2)^2}(\ell -1)(\ell +2) \times \nonumber\\
&&\!\!\!\!\!\!\!\!\!\!\!\!
\left( {(\ell+m+1)(\ell -m+1)(\ell+m)(\ell -m) \over
 (2\ell-1) (2\ell +1)^2(2\ell +3)}\right)^{1/2}
{\Ga(\ell+n/2+3/2)\over \Ga(\ell-n/2+1/2)}~.
\eea
This result is valid in the range $-7<n<-1$. For $n \le-7$ the quadrupole
diverges at small k, and for $n>-1$ the result is dominated by the
upper cutoff $k_0$,
\be
C_\ell \simeq D_\ell \simeq
{v_A^2A_0 \over 2 \pi 
(n+1)(k_0 t_0)^2 } \left( {t_{dec} \over t_0} \right)^2\ell^2
~~~~,~~~~~~~~ n>-1~.
\ee
 For $n=-5$ we obtain a scale-invariant
Harrison-Zeldovich spectrum, $C_\ell\sim \ell^2$.  
 
To obtain some insight into the effect of the cross terms 
$D_\ell$, we picture the correlation function
\be
f({\bf n})= \left\langle{\delta T\over T}({\bf n_0}){\delta T\over T}({\bf n})
\right\rangle\label{f_eqn}
=
 \sum_{\ell m \, \ell' m'} \langle a^{}_{\ell m}a^*_{\ell' m'}\rangle 
	Y_{\ell m}(\bn_0)Y^*_{\ell' m'}(\bn)
\ee
for various orientations of the magnetic field with respect to the
fixed direction ${\bf n_0}$. These are shown for the case 
$n=-5$ and with ${\bf n_0} = \hat {\bf z}$ in
figures~1 to 3. Notice, however, that these figures 
do not represent temperature maps but are plots of the correlation
function. For a given realization stochastic noise has to be added.
The explicit expression for $f(\bn)$ is given in Appendix~A.

With no {\it a priori} knowledge of the
field direction, it could be inferred by performing CMB measurements
with various ${\bf n_0}$ and comparing the obtained $f({\bf n})$ with
the plots below. Of course this procedure suffers from problems with
cosmic variance, as once we fix a direction ${\bf n_0}$ in the sky we
have only a single realization with which to determine $f$.
The expectation value in expression (\ref{f_eqn}), then, strictly
refers to a (hypothetical) average over an ensemble of universes.
 
\begin{figure}
\centerline{\epsfxsize=5in\epsfbox{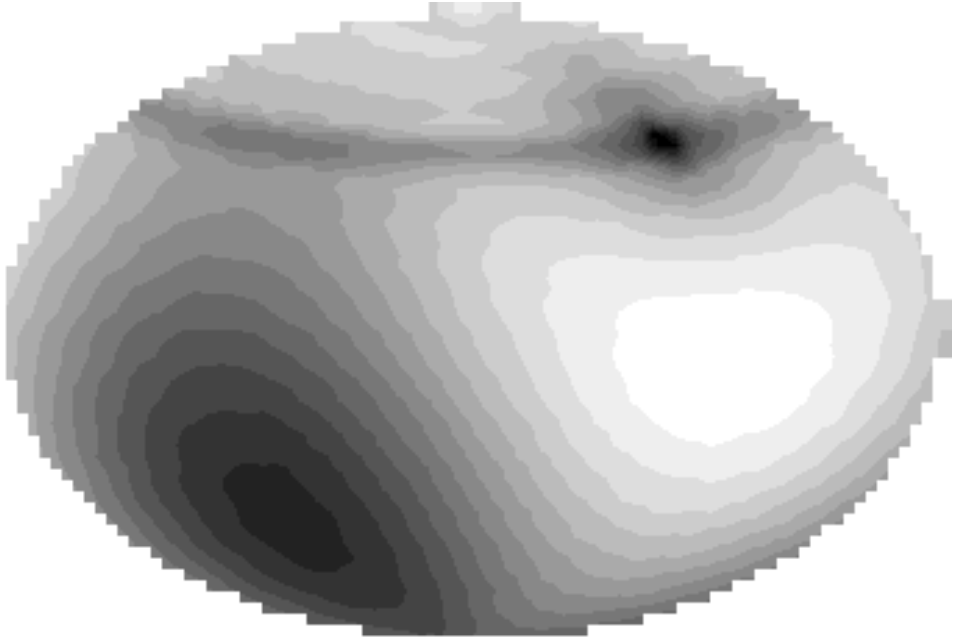}}
\caption{An Aitoff  projection of the function $f(\bn)$ for a homogeneous
magnetic field pointing in the $\theta=\pi/4$, $\phi=\pi/2$ direction
and the reference 
vector $\bn_0$ pointing in the $z$-direction ($\theta=0)$
(see equation (\ref{f_eqn})).}
\label{map1}
\end{figure}
\begin{figure}
\centerline{\epsfxsize=5in\epsfbox{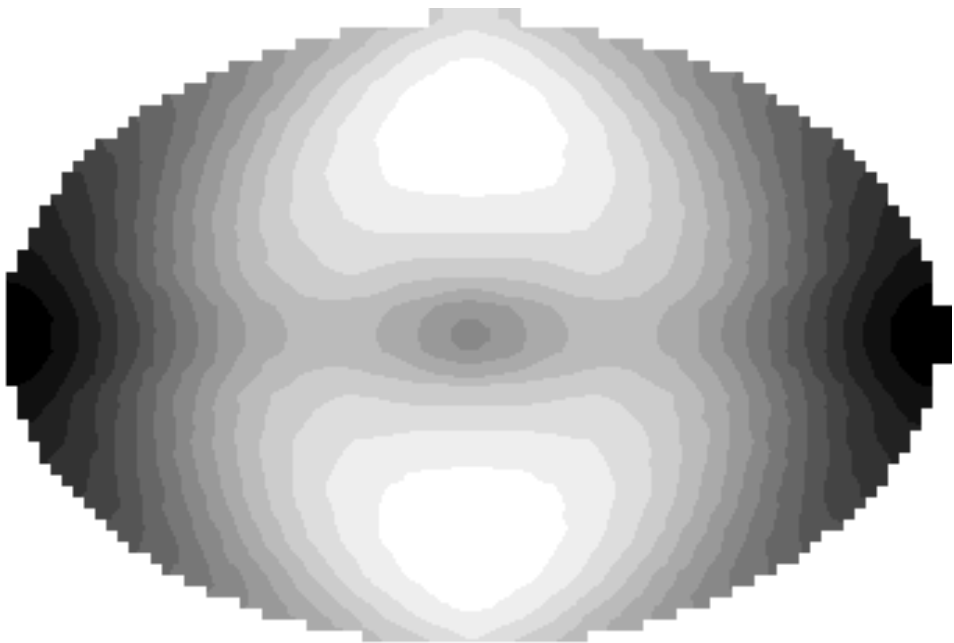}}
\caption{The function $f(\bn)$ for ${\bf B_0}$ pointing in the
$\theta=\pi/2$, $\phi=0$ direction.}
\label{map2}
\end{figure}
\begin{figure}
\centerline{\epsfxsize=5in\epsfbox{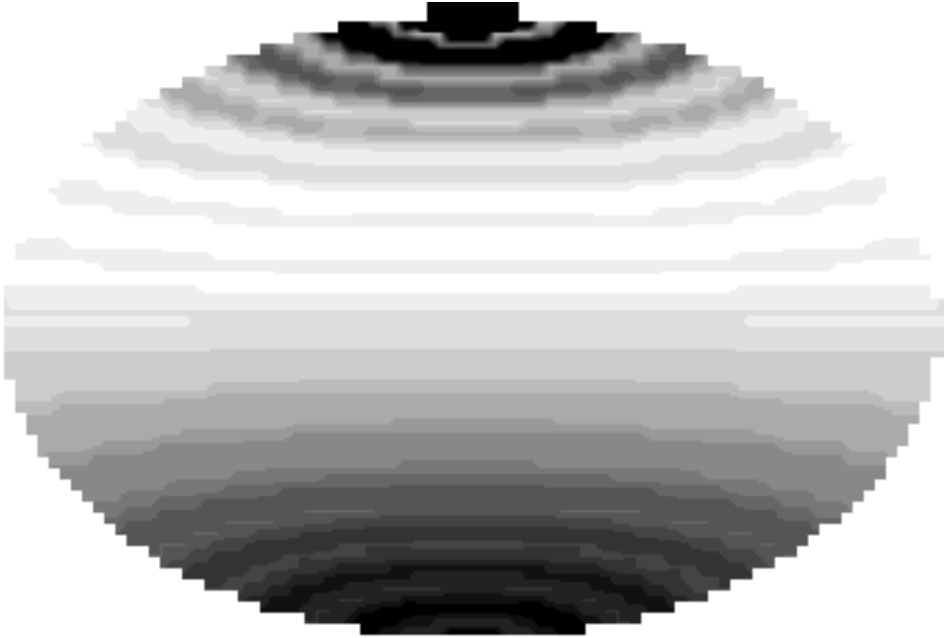}}
\caption{The function $f(\bn)$ for  ${\bf B_0}$ pointing in the
$\theta=0$ direction ({\it i.e.}, parallel to $\bn_0$). The gray scale
scheme has enhanced the variation in $f$.}
\label{map3}
\end{figure}

A probably simpler observational test of the 
existence of a constant magnetic field is the presence of
temperature correlations for unequal $\ell$. To simplify, we introduce the
mean values 
\bea
\overline{C}_\ell &=& \overline{\langle a^{}_{\ell m}a^*_{\ell
m}\rangle}~,\\
\overline{D}_\ell &=& \overline{\langle a^{}_{\ell-1, m}a^*_{\ell+1,
m}\rangle}~, 
\eea
where the bar denotes average over different values of $m$, and we find
\bea
\overline{C}_\ell &\simeq& A_0\left({t_{dec}\over
t_0}\right)^2(k_0t_0)^{-(n+3)}v_A^2{2^{n+1}\Ga(-n-1)\over
3\Ga(-n/2)^2}\ell^{n+3}  , ~~~\mbox{ for } n<-1 \label{Cbar}\\ 
\overline{C}_\ell &\simeq& A_0\left({t_{dec}\over
t_0}\right)^2(k_0t_0)^{-2}v_A^2{1\over n+1}\ell^2 ,
	 ~~~\mbox{ for } n>-1 \label{Cbar+}\\ 
 \overline{D}_\ell/ \overline{C}_\ell &\simeq& 3/2 ~.\label{Dbar}
\eea
The existence of significant correlations between the $a_{\ell-1, m}$ and
 $a_{\ell+1, m}$ is a clear indication of the presence of a preferred
direction in the universe. Due to its spin-1 nature, a long-range
vector field induces 
transitions $\ell\ra\ell\pm 1$ and thus leads to the correlators
$D_{\ell}$.

There are no published limits on these cross correlation terms. Since
the full $a_{\ell m}$'s are needed to obtain them, full sky coverage
and high resolution, as will be provided by the MAP and PLANCK
satellites, are most important to obtain such limits. The galaxy cut
in the 4-year COBE data leads to an influence of $C_\ell$ by
$C_{\ell\pm 2}$ which is on the order of 10\%. for $2 \le \ell \le 30$
\cite{Wright}. To be specific, let us assume that this be a limit on
the off diagonal correlations $D_{\ell}$. Then, first of all,
the observed CMB anisotropies 
are not due to Alfv\'en waves, since  $0.1 \simeq D_\ell /
C_\ell$ is substantially smaller than the figure in Eq.~(\ref{Dbar}). 
To obtain a limit on the magnetic field amplitude and the spectral 
index, we now require 
\be
\ell^2{\overline{D}_\ell}\le 0.1 \ell^2{\overline{C}_\ell}\simeq
	10^{-11} ~~\mbox{ for }~~ 2 < \ell\lsim 100~. \label{-10}
\ee
We now argue as follows. From Eq.~(\ref{amp}), and the fact that ${\bf
B}_1\lsim{\bf B}_0$, we have
\be
|\bm{\Om}_0|^2k^3\lsim v_A^2 ~.
\ee
This inequality must hold on all scales inside the horizon at
decoupling, $k\ge 1/t_{dec}$. With Eq.~(\ref{A_spec}) we therefore obtain
\be
 2A_0(k/k_0)^{n+3}\lsim v_A^2 \qquad 1/t_{dec} \le k \le k_0, 
\ee
which implies
\bea
 2A_0(k_0t_{dec})^{-(n+3)} &\lsim& v_A^2 ~~\mbox{ for }~ n\le -3
	\label{lim1},\\
 2A_0 &\lsim& v_A^2 ~~\mbox{ for }~ n\ge -3 ~.\label{lim2}
\eea
Here we have identified $k_0$ with the maximal frequency (cutoff) of
the magnetic field, which has to be introduced in the case $n>-3$ for
$\bm{\Om}$ not to diverge at small scales.  A definite upper limit on
$k_0$ is the 
scale beyond which the magnetic field is damped away, due to the
finite value of the conductivity.
The physical damping scale is given by \cite{Jackson} 
\be
	(k_D/a)^2 = 4\pi\si/\tau~,
\ee
where $\tau$ denotes the cosmic time (not comoving) and $\si$ is the
plasma conductivity. The conductivity
of a non-relativistic electron-proton plasma is easily shown to
be  $\si\sim 4T$, and it has been shown recently that this result 
still holds approximately in the very early universe~\cite{AhEn}.

Using $T_{dec}\sim 0.3\,{\rm eV} \sim 0.6 \times 10^{-4}$cm$^{-1}$ and 
$\tau_{dec}\sim 10^5\,$years~$\sim 10^{23}\,$cm, we obtain 
the comoving damping scale at decoupling 
\be
k_0(t_{dec}) \sim k_D(t_{dec}) \sim (z_{dec})^{-1}\sqrt{
16\pi T_{dec}/\tau_{dec}} \sim 3\times 10^{-10}\mbox{cm}^{-1},
\ee
and
\be
(k_0t_0)(t_{dec}) = k_0(t_{dec})\tau_0/a_{dec} \sim
\tau_0 \sqrt{16\pi T_{dec}/\tau_{dec}}\sim 0.4\times 10^{14}.
\ee
Inserting the limiting values of Eqs.~(\ref{lim1}, \ref{lim2}) for the
$A_0$ in Eq.~(\ref{Cbar}), Eq.~(\ref{-10}) yields
\bea
3v_A^4z_{dec}^{-(n+5)/2}{2^{n+1}\Ga(-n-1)\over
3\Ga(-n/2)^2}\ell^{n+5} < 10^{-11} & \mbox{ for} & n<-3, \label{lim3-}\\
3v_A^4z_{dec}^{-1}(2.5\times 10^{-14})^{(n+3)}{2^{n+1}\Ga(-n-1)\over
3\Ga(-n/2)^2}\ell^{n+5} < 10^{-11} & \mbox{ for} &
	 -3\le n < -1,\label{lim1-}\\
3v_A^4z_{dec}^{-1}(2.5\times 10^{-14})^2{1\over n+1}\ell^4 < 10^{-11} 
& \mbox{ for} & -1<n, \label{lim1+}
\eea
Using $v_A\sim 4\times 10^{-4}(B_0/10^{-9}\mbox{Gauss})$, this can be
translated into a limit for $B_0$ which depends on the spectral index
$n$ and the harmonic $\ell$. In Fig.~4 we plot the best limit on $B_0$
as a function of the spectral index $n$. To optimize the limit we
choose $\ell=2$ for $n<-5$ and $\ell=100$ for $n>-5$. For $n>-3$ the
limit becomes very quickly entirely irrelevant due to the huge factor
$ 10^{14(n+3)}$. This reflects the fact that  for $n>-3$, the
magnetic field fluctuations grow towards small scales, and $B_1\lsim B_0$ is 
leads to a limit at the tiny scale $k=k_0$; where as the CMB anisotropies,
 are caused by the smaller fluctuations at large scales,
$k\sim \ell/t_0$. At $n\le -1$ the induced $D_\ell$'s start to feel the upper
cutoff and thus do not decrease any further. 
\begin{figure}
\centerline{\epsfxsize=10cm
\epsfbox{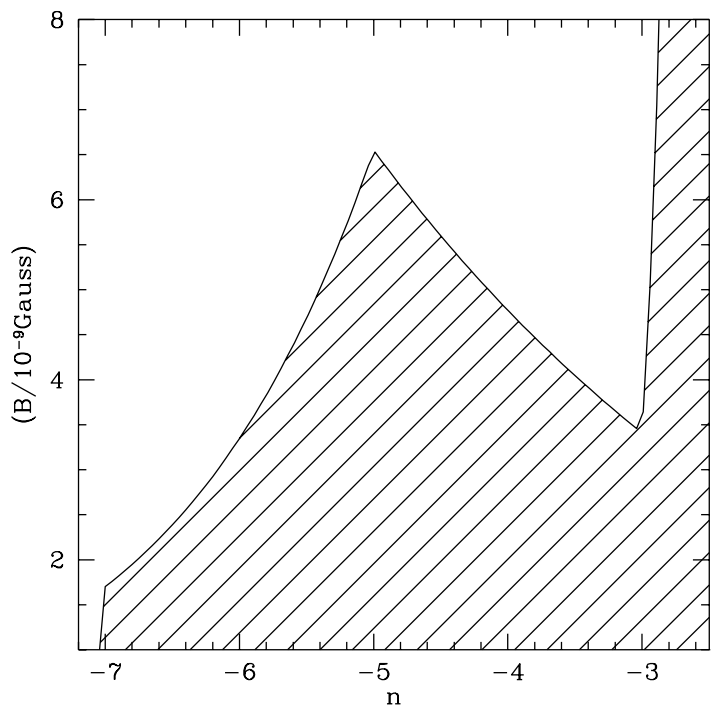}}
\caption{The upper limit on the magnetic field amplitude $B_0$ due to CMB 
anisotropies caused by Alfv\'en waves, shown as a function of the
magnetic field spectral index $n$. We assume $D_\ell \le 0.1
C_\ell$. Allowed values of the field must lie in the dashed region.}
\label{fig4}
\end{figure}
 
The presence of a homogeneous magnetic field also induces
anisotropic stresses in the metric. This gravitational effect has 
been estimated elsewhere \cite{BFS}. Compared with the 
COBE DMR  experiment, it leads to a similar limit for $B_0$.

\section{Conclusions}
We have studied Alfv\'en waves in the primordial electron-proton
plasma that are sourced by a homogeneous magnetic field.  
In addition, we allow for an isotropic
distribution of random magnetic fields on smaller scales.
The induced vorticity in the baryon fluid leads, via the Doppler effect,
to vector-type CMB anisotropies on all angular scales larger than 
the vectorial Silk damping scale $\ell_{damp}\sim 500$ (see Appendix~B).
The vector nature of the magnetic field induces off-diagonal
correlations, 
\be
D_\ell(m) =\langle a_{\ell-1, m} a^*_{\ell+1, m}\rangle \sim C_\ell(m) ~.
\ee
Assuming that observations constrain these terms to be less than about
10\% of the observed $C_{\ell}$'s, we derive a limit for
the amplitude of the 
magnetic field as a function of its spectral index. For $n<-7$, the
quadrupole anisotropy diverges if no lower cutoff is imposed on the
spectrum, and so such spectra are very strongly constrained. For $n>-3$, the
constraint is proportional to $(k_0t_0)^{(n+3)/4}$, where $t_0$ is the
comoving scale today and $k_0$ is the upper cutoff of the spectrum. We
have set $1/k_0$ equal to the magneto-hydrodynamical damping scale which
is inversely proportional to the conductivity and thus extremely small,
leading to $(k_0t_0)\sim 10^{14}$. Therefore, the limits obtained for
$n>-3$ are extremely weak and actually uninteresting. This is due to
the fact that the quantity $B^2(k)k^3$ decreases on large scales 
 for $n>-3$. For spectral indices in the
range $-7<n<-3$ the limit on $B_0$ is on the order of
$(2-7)\times10^{-9}$Gauss. 

An important remark is also that causally induced magnetic fields
lead to a spectral index $n=2$ and so are not constrained at
all\footnote{If we assume magnetic fields to be generated by a   
causal procedure, {\em i.e.} not during an inflationary epoch, in a
Friedmann universe, then the real space correlation function has to 
vanish at super-horizon distances (say $|x|>2t$). Its Fourier transform 
$\langle B_i B_j\rangle({\bf k})\propto k^n(\de_{ij}-k_ik_j)$ is therefore
analytic in $\bk$, which requires $n$ to be an integer with $n\ge 2$.}.  
Examples here are magnetic fields generated by the decay of a $Y$
field during the electroweak phase transition.

At first sight it may seem somewhat artificial to have split the
magnetic field into a homogeneous component and an isotropic spectrum of
random magnetic fields. However, this is the correct procedure for
realistic observations. This is seen as follows. If we calculate the
$C_\ell$'s for a given model, we determine expectation values
over an ensemble of universes. If we make a measurement,
however, we have just one observable universe at our disposition.
This problem is generally referred to as `cosmic variance'. On
scales much smaller than the horizon, cosmic variance is
irrelevant if we make some kind of ergodicity hypothesis, assuming
that spatial and ensemble averaging are equivalent. In the case
here cosmic variance is especially important, however. If we include
the field which is coherent on the horizon scale in our random
distribution, then in performing the
ensemble average we integrate over the directions of ${\bf B}_0$ and the
off-diagonal correlators $D_\ell$ vanish. The entire effect
disappears. In one given universe, however, this field 
has one fixed direction and our effect is observable. It
is therefore of fundamental importance here not to take an ensemble
average over the large scale coherent field.
The cosmic variance problem 
is relevant, whenever perturbations with non-vanishing power on
horizon scales are present. 

An observational limit on the off diagonal
correlators $D_\ell$'s  from MAP or PLANCK would represent a
model independent limit on the importance of large scale coherent
vector fields (which enter quadratically in the energy momentum
tensor) for the  anisotropies in the cosmic microwave background.
Its importance thus goes beyond the magnetic field case discussed in
the present work.
\vspace{0.3cm}\\
{\bf \large\bf Acknowledgments}: It is a pleasure to acknowledge
discussions with Kris Gorski, Alessandro Melchiorri and Misha
Shaposhnikov. Francesco Melchiorri brought our attention to reference
\cite{Wright}.  This work is financially supported by the Swiss
NSF. T.K. is grateful for hospitality at Geneva University.

\vspace{3cm}

\begin{appendix}

\section{Calculation of $C_{\ell}$ from vector perturbations}
In the usual way we decompose the temperature fluctuations of the
microwave background into spherical harmonics:
\[
{\delta T \over T} (\bn) = \sum_{\ell m} a_{\ell m} Y_{\ell m}(\bn).
\]
The two point function $\langle a^{}_{\ell m} a^*_{\ell' m'}
\rangle$ is then
\be
\langle a^{}_{\ell m} a^*_{\ell' m'} \rangle
= {1\over (2\pi)^3} \int d^3k \int d\Om_{\bn} \int d\Om_{\bn'}
\left\langle {\delta T\over T}^* (\bn,\bk){\delta T\over T} (\bn ',\bk)
\right \rangle
Y^*_{\ell m} (\bn) Y_{\ell 'm'}(\bn ')~.
\ee
We consider the contribution to the temperature anisotropy only
from the vorticity in the baryon fluid. From Eq.~(\ref{tb}), we
obtain
\be
{\delta T\over T} (\bn, \bk, \De t) = e^{i \bk \cdot \bn \De t} \bn
\cdot {\bf 
\Om}({\bf  
k}, t_{dec})
\label{det} ~,
\ee
where $\De t=t_0 - t_{dec} \simeq t_0$, the time elapsed since last
scattering. Using the form (\ref{spec}) for the power spectrum of
the vortical velocity fluctuations, we have
\bea 
\left \langle {\delta T\over T}^* (\bn,\bk){\delta T\over T} (\bn ',\bk)
\right \rangle
& = & e^{i\bk\cdot (\bn' - \bn)t_0} (\bn\cdot \bn' - \mu \mu') (v_A k 
\beta t_{dec})^2 A(k) \nonumber \\
&=&
  A(k) (v_A k \beta)^2 
\left( e^{ikt_0(\mu' - \mu)}\bn\cdot \bn' -  {\dd\over \dd (kt_0)}
e^{-ikt_0\mu} {\dd\over \dd (kt_0)} e^{ikt_0\mu'} \right ) \nonumber \\
& = & f(\bn, \bn', \bk),
\eea
where $\mu = \bn\cdot \hat \bk $, $\mu' = \bn' \cdot \hat \bk $, 
$\beta = {\bf e} \cdot \hat \bk $ and ${\bf e}$ is the unit vector in
the direction of the homogeneous magnetic field ${\bf B}_0$. 
So 
\be
\langle a^{}_{\ell m} a^*_{\ell' m'} \rangle
= {1\over (2\pi)^3} \int d^3k \int d\Om_\bn \int d\Om_{\bn'} \; f(\bn,
\bn', \bk)\; 
Y^*_{\ell m} (\bn) Y^{}_{\ell'm'}(\bn') ~.\label{almalm}
\ee
To evaluate these integrals, we use the identities
\bea
e^{ix \hat \bk\cdot \bn} &=& 4\pi \sum_{r=0}^{\infty} \sum_{q=-r}^{+r}
i^r j_r(x) 
Y_{rq}^* (\hat \bk) Y^{}_{rq}(\bn) \label{expexp}~,\\
 \bn \cdot \bn' & =&  P_1(\bn\cdot\bn') = {4\pi \over 3}
\sum_{p=-1}^{+1} Y_{1p}(\bn) Y_{1p}^* (\bn') \label{ndotn}~,
\eea
where $j_r$ is the spherical Bessel function of order $r$. Using the
orthonormality of the spherical harmonics and the recursion relation
\be
(2\ell +1) j_{\ell}' = \ell j_{\ell -1} -
(\ell+1) j_{\ell+1}~,
\ee
we find that in evaluating (\ref{almalm}) only the terms with
$(\ell,m)=(\ell',m')$ and $(\ell,m)=(\ell'\pm 2,m')$ survive, where
\bea
\langle a^{}_{\ell m} a_{\ell m}^* \rangle &=&
\left( 
{2\ell^4 + 4\ell^3 - \ell^2 - 3\ell + 6m^2 - 2\ell m^2 - 2\ell^2 m^2
\over (2\ell-1) (2\ell+1)^2 (2\ell+3) } 
\right)
\times \nonumber \\
&& {2\over \pi} \int dk\,k^2 \, (v_A k t_{dec})^2A(k)
	(j_{\ell+1} + j_{\ell -1} )^2
\eea
and
\bea
\lefteqn{\langle a^{}_{\ell+1 m} a^*_{\ell-1, m} \rangle =
\langle a^{}_{\ell-1 m} a^*_{\ell+1, m} \rangle = } \nonumber \\
&& -(\ell-1)(\ell+2)
\left( {(\ell+m+1)(\ell-m+1)(\ell+m)(\ell-m)
\over
(2\ell-1)^3(2\ell+1)^2(2\ell+3)^3}
\right)^{1/2}\times
  \nonumber \\
&&{2\over \pi} \int dk\, k^4\, (v_A t_{dec})^2  A(k)\left(
 j_{\ell} + j_{\ell-2} \right) \left( j_{\ell} + j_{\ell+2} \right).  
\eea 
The Bessel functions take $kt_0$ as their arguments. With $A(k)=
A_0(k/k_0)^n k_0^{-3}$, we obtain, for $-7 < n < -1$,  
\bea
\langle a^{}_{\ell m}a^*_{\ell m}\rangle &\equiv& C_\ell(m) \nonumber\\
&& {2^{n+1} A_0v_A^2 \over  (k_0 t_0)^{n+3}}
\left( {  t_{dec} \over t_0} \right)^2 
{\Gamma(-n-1) \over \Gamma(-n/2)^2}
{\Gamma(\ell + n/2 + 3/2) \over \Gamma(\ell -n/2 +1/2)}\times
\nonumber \\
&& { (2\ell^4 + 4\ell^3 - \ell^2 - 3\ell + 6m^2 - 2\ell m^2 - 
2\ell^2 m^2) \over (2\ell-1)(2\ell+3)},  \\
\langle a^{}_{\ell+1, m}a^*_{\ell-1, m}\rangle &=& 
\langle a^{}_{\ell-1, m}a^*_{\ell+1, m}\rangle \equiv D_\ell(m)\nonumber\\
&&
{2^{n+2}A_0 v_A^2  \over |n+1| (k_0t_0)^{n+3}}
\left( {t_{dec} \over t_0} \right)^2 
{ \Gamma(-n-1) \over \Gamma(-(n+1)/2)^2}
{\Gamma(\ell + n/2 + 3/2) \over \Gamma(\ell - n/2 + 1/2) }\times
\nonumber \\
&&\!\!\!\!\!\!\!\!
{(\ell -1)(\ell +2)}
\left( {(\ell+m+1)(\ell -m+1)(\ell+m)(\ell -m) \over
 (2\ell-1) (2\ell +1)^2(2\ell +3)}\right)^{1/2}.
\eea
For $n>-1$ the integral is dominated by the upper cutoff $k_0$ and we
find
\bea
C_\ell(m) &=& 
{v_A^2 A_0\over 2 \pi 
(n+1)(k_0 t_0)^2 } \left( {t_{dec} \over t_0} \right)^2 \times\\
&& { (2\ell^4 + 4\ell^3 - \ell^2 - 3\ell + 6m^2 - 2\ell m^2 - 
2\ell^2 m^2) \over (2\ell-1)(2\ell+3)},  \\
D_\ell(m) &=&{v_A^2 A_0\over 2 \pi 
(n+1)(k_0 t_0)^2 } \left( {t_{dec} \over t_0} \right)^2\times \\
&&\!\!\!\!\!\!\!\!
{(\ell -1)(\ell +2)}
\left( {(\ell+m+1)(\ell -m+1)(\ell+m)(\ell -m) \over
 (2\ell-1) (2\ell +1)^2(2\ell +3)}\right)^{1/2}~.
\eea
In this case, the result is nearly
independent of the spectral index $n$ and, due to the factor
$(k_0t_0)^{-2}$,  it is so small that it 
fails to lead to relevant constraints for $B_0$.

The temperature correlation function is finally
\bean
f({\bf n}) & = &
\left\langle{\delta T\over T}({\bf n_0}){\delta T\over T}({\bf n})
\right\rangle\label{f_eqn2}
=
\sum_{\ell m \, \ell' m'} \langle a^{}_{\ell m}a^*_{\ell' m'}\rangle 
	Y^{}_{\ell m}(\bn_0) Y^*_{\ell' m'}(\bn)\\
&=& \sum_{\ell m} \langle a^{}_{\ell m}a^*_{\ell m}\rangle
Y^{}_{\ell m}(\bn_0)Y^*_{\ell m}(\bn) +\\
&& \sum_{\ell m} \langle a^{}_{\ell+1,m}a^*_{\ell-1,m}\rangle
\left(
Y^{}_{\ell+1, m}(\bn_0)Y^*_{\ell-1, m}(\bn) +
Y^{}_{\ell-1, m}(\bn_0)Y^*_{\ell+1, m}(\bn)
\right). 
\eean

\section{Collisional damping for vector perturbations}
Denoting the fractional perturbation in the radiation brightness  by
$\MM$, $\MM=4(\De T/T)$, the Boltzmann equation for vector
perturbations gives \cite{fc}
\be
 	\dot\MM +\bn\cdot \nabla\MM = -4n^in^j\si_i,_j   + 
	a\si_Tn_e\left[-\MM+4\bn\cdot\bm{\Om}\right] ~. \label{coll}
\ee
Here $\bn$ is the photon direction, $\si_T$ denotes the Thomson cross
section  and $\bm{\Om}$ is the baryon vorticity. We have neglected the
anisotropy of non-relativistic Compton scattering. 

To the baryon equation of motion (\ref{cons}) we have to add
the photon drag force,
\be
	\dot{\bm{\Om}}+ {\dot{a}\over a}\bm{\Om}={a\si_Tn_e\rho_r\over
	3\rho_b}\left[{1\over 4}{\bf M}  - \bm{\Om}\right]~,
\label{drag} 
\ee
with
\[
{\bf M} ={3\over 4\pi}\int\bn\MM d\bn ~.
\]
We shall also use the fact that for vector perturbations, the
perturbation of the photon brightness vanishes,
\[
\int\MM d\bn =0~.
\]
Due to the loss of free electrons during recombination, the mean 
(conformal) collision time $t_c=1/(a\si_Tn_e)$ increases from 
a microscopically small scale before recombination to a super-horizon
scale after recombination. After recombination the collision term can
be neglected and we recover Eqs.~(\ref{cons}) and (\ref{Tv}).
We first consider the very tight coupling regime, $t_c\ll \la$,
where $\la$ denotes the typical scale of fluctuations. In this
limit the term inside the square brackets of Eqs. (\ref{coll}) and
(\ref{drag}) can be set to zero and we obtain ${\bf M} =4\bm{\Om}$ (
the baryon and photon fluids are adiabatically coupled). 

Next, we derive a dispersion relation for the damping of fluctuations
due to the finite size of $t_c$. We proceed in the same way as
Peebles \cite{Pee80} for scalar perturbations. We consider scales with
wavelength $k^{-1}\ll t$ and thus neglect the time dependence of the
coefficients in Eqs.~(\ref{coll}) and (\ref{drag}). To study the damping we
also neglect gravitational effects, which act on much slower
timescales. With the ansatz 
\bea
	\MM &=& \A(\bn)\exp(i(\bk\bx-\om t)),  \\
	\bm{\Om} &=& \B(\bn)\exp(i(\bk\bx-\om t)), \;\;\;(\B\cd\bk)=0
\eea
we obtain
\bea
 -i\om\A +i(\bk\bn)\A &=& {1\over t_c}[-\A+4\bn\cdot \B]   \\
 -i\om\B &=& {1\over t_cR}\left[{\bf M} -4\B\right]~,~~~ R \equiv
        {3\rho_b\over 4\rho_\ga}~. \label{Reqn}
\eea
In the limit $kt_c,\om t_c\ra 0$, we again obtain adiabatic coupling.
The general relation between $\A$ and $\B$ is
\[
\A={4\bn\cdot\B\over 1 +i(\bk\cdot\bn-\om)t_c}
\]
and so
\be
{\bf M} =3\B{i\over (kt_c)^3}\left[
 	-((kt_c)^2+(1-i\om t_c)^2)\ln\left({1-it_c(\om-k)\over 
	1-it_c(\om+k)} \right) +2ikt_c(1-i\om t_c)\right]~. \label{int}
\ee
Inserting this is in (\ref{Reqn}) leads again,
in the limit $kt_c,\om t_c\ra 0$, to the tight coupling result. In
first order $kt_c$ (the square 
bracket in Eq.~(\ref{int}) has to be expanded up to order $(kt_c)^5$)
we obtain the dispersion relation 
\be \om =-i\ga, ~~\mbox{ with }~ \ga={7k^2t_c\over 20(1+R)} \sim
{k^2t_c \over 3}~.\label{disp}\ee
In contrast to the scalar case, vector perturbations show no
oscillations (${\rm Re}(\om) =0$) but are just damped. The damping occurs
at a slightly larger scale than for scalar perturbations, where
$\ga_{scalar}\simeq k^2t_c/6$ \cite{Pee80}.

The ratio $R=3\rho_b/(4\rho_\ga)$ is smaller than $\sim 1/4$ until
the end of recombination. We therefore obtain a damping factor $f$ for a 
given scale $k$ 
\be
f \sim \exp\left( {7k^2\over 20}\int_0^{t_{end}(k)}t_cdt \right)
~,\label{damp}
\ee 
where $t_{end}(k)$ is the time at which our approximation $kt_c<1$
breaks down, {\em i.e.}, $kt_c(t_{end}(k))=1$. The time over which the
damping is active is the order of the thickness of the last
scattering surface, $\De t\sim t_{dec}(\De z/z_{dec})\sim
0.1t_{dec}$.
The damping scale, the scale at which the exponent in Eq.~(\ref{damp})
becomes of order unity, is about
\be
k_{damp}t_{dec} \sim 10.  \label{kdamp}
\ee 
The harmonic $\ell$ corresponding to $k_{damp}$ is $\ell_{damp} =
k_{damp} t_0 \sim 10 t_0/t_{dec} \sim 500$.

After the time $t_{end}(k)$, collisions become unimportant for
fluctuations with wave number $k$ which then evolve freely, suffering
only directional dispersion which induces a power law
damping $\propto 1/(k\De t)$. Reference \cite{Barrow} discusses only
this second effect. Numerical experience with scalar perturbations,
however, shows that they are typically both of similar importance.

\end{appendix}
\end{document}